\documentclass{aa}

\usepackage{natbib}
\bibpunct{(}{)}{;}{a}{}{,} 

\usepackage{float}
\usepackage{graphicx}
\usepackage{placeins}
\usepackage{caption}
\DeclareCaptionFormat{cont}{#1 (cont.)#2#3\par} 
\usepackage{txfonts}
\usepackage{siunitx}  
\usepackage{color}
\usepackage{amsmath}
\usepackage{dblfloatfix}    
\usepackage[flushleft]{threeparttable} 
\usepackage{ulem}

\usepackage[title]{appendix}

\begin{document}

\title{The first detection of SiC$_2$ in the interstellar medium}

\titlerunning{}
\authorrunning{Massalkhi et al.}
\author{S. Massalkhi$^1$, I. Jim\'enez-Serra$^1$, J. Mart\'in-Pintado$^1$, V. M. Rivilla$^{1}$, L. Colzi$^{1}$, \\ S. Zeng$^2$, S. Mart\'in$^{3,4}$, B. Tercero$^5$, P. de Vicente$^5$, and M.A. Requena-Torres$^{6}$}

\institute{$^1$ Centro de Astrobiología (CAB), INTA-CSIC, Carretera de Ajalvir km 4, Torrej\'on de Ardoz, 28850, Madrid, Spain.\\ 
$^2$ Star and Planet Formation Laboratory, Cluster for Pioneering Research, Riken, 2-1 Hirosawa, Wako, Saitama, 351-0198, Japan. \\
$^3$ European Southern Observatory, Alonso de C\'ordova, 3107, Vitacura, Santiago, 763-0355, Chile.\\
$^4$ Joint ALMA Observatory, Alonso de C\'ordova 3107, Vitacura 763 0355, Santiago, Chile.\\
$^5$ Observatorio de Yebes (IGN), Cerro de la Palera s/n, 19141, Guadalajara, Spain.\\
$^6$ Department of Physics, Astronomy and Geosciences, Towson University, 8000 York Road, Towson, MD 21252, USA.}
\date{Received; accepted}

\abstract{We report the first detection of SiC$_2$ in the interstellar medium. The molecule was identified through six rotational transitions toward G\,+0.693$-$0.027, a molecular cloud located in the Galactic center. The detection is based on a line survey carried out with the GBT, the Yebes 40m, and the IRAM 30m telescopes covering a range of frequencies from 12 to 276 GHz. We fit the observed spectra assuming local thermodynamic equilibrium and derive a column density of ($1.02\pm0.04)\times10^{13}$ cm$^{-2}$, which gives a fractional abundance of $7.5\times10^{-11}$ with respect to H$_2$, and an excitation temperature of $5.9\pm0.2$ K. We conclude that SiC$_2$ can be formed in the shocked gas by a reaction between the sputtered atomic silicon and C$_2$H$_2$, or it can be released directly from the dust grains due to disruption. We also search for other Si-bearing molecules and detect eight rotational transitions of SiS and four transitions of Si$^{18}$O. The derived fractional abundances are $3.9\times10^{-10}$ and $2.1\times10^{-11}$, respectively. All Si-bearing species toward G\,+0.693$-$0.027 show fractional abundances well below what is typically found in late-type evolved stars.}


\maketitle

\section{Introduction}

In the envelopes around asymptotic giant branch (AGB) stars, gas-phase silicon-bearing molecules, such as SiO, SiS, and SiC$_2$, are formed efficiently under local thermodynamic equilibrium (LTE) close to the stellar photosphere \citep{tsu1973}. These molecules are then injected into the expanding envelope where they are likely to deplete from the gas phase and condense onto dust grains due to their highly refractory character. Their abundance will therefore experience a decline due to adsorption onto grains. New Si-bearing molecules may also be formed via grain-surface chemistry (e.g., \citealt{cer2017}). Moreover, periodic shock waves caused by  stellar pulsations can propagate through the photosphere, releasing Si-bearing molecules and/or atomic Si from the dust grains, further altering the silicon chemistry (e.g., \citealt{cas2001}). Photochemistry may also give rise to other Si-bearing molecules in the external shells (e.g., \citealt{cer1989}). In the outermost regions of the envelope, the ambient interstellar ultraviolet field will destroy the free gas-phase molecules through photodissociation, whereas the dust may survive and be incorporated into the interstellar medium (ISM).

The situation is different in the ISM, where molecules containing silicon have been extremely elusive. Of the 13 Si-bearing molecules currently detected toward evolved stars, only 3 (SiO, SiS, SiN) are observed in the ISM \citep{wil1971,mor1975,sch2003}.  SiC$_2$ \citep{tha1984} is an interesting molecule because it is believed to be a precursor of SiC dust. The presence of SiC grains in carbon evolved stars has been identified through the solid-state emission feature at $\sim$\SI{11.3}{\micro\meter} \citep{hac1972}. \citet{mas2018} carried out observations toward a sample of 25 carbon-rich AGB stars and observed a trend whereby the denser the envelope, the lower the abundance of SiC$_2$. This was interpreted as evidence of efficient adsorption of SiC$_2$ onto dust grains. However, it is remarkable that this molecule has not yet been observed in the interstellar space, especially given that SiC dust is ubiquitous in carbon-rich evolved stars and it is through their stellar outflows that the dust is transported to the ISM. 

Here, we report the discovery of SiC$_2$ in the ISM. The detection was made toward the molecular cloud G\,+0.693$-$0.027 (hereafter G\,+0.693) located in the Galactic center (GC) within the Sgr B2 complex. The cloud is known to be chemically rich \citep{req2006,zen2018,jim2020} and its chemistry is believed to be the result of a large-scale cloud--cloud collision \citep{zen2020}. The collision drives low-velocity shocks ($\sim$20 km s$^{-1}$), which sputter the molecular content of the grain icy mantles into the gas phase (e.g., \citealt{mar2008,jim2008}). A series of complex molecules was recently discovered in G\,+0.693, such as hydroxylamine, ethanolamine, vinyl amine, ethyl isocyanate, monothioformic acid, and n-propanol \citep{riv2019,riv2020,riv2021b,riv2022a,zen2021,rod2021a,rod2021b,jim2022}, as well as other simpler ones, like HNCN and PO$^+$ \citep{riv2021a,riv2022b}, making this cloud an excellent target in which to search for new molecular species. 

The paper is organized as follows. In Sect.~\ref{section:observations}, we provide details of the observations. In Sect.~\ref{section:results}, we describe the LTE and nonLTE analysis and present the results from these calculations.  In Sect.~\ref{section:discussion}, we discuss the main findings of our study, and finally we lay out our conclusions in Sect.~\ref{section:conclusion}. 

\section{Observations} \label{section:observations}
The observational data used in this article are based on a spectral survey carried out with the Green Bank Telescope (GBT) in West Virginia, USA, the Yebes 40m telescope in Guadalajara, Spain, and the IRAM 30m telescope in Pico Veleta, Spain. The equatorial coordinates of the target source, G\,+0.693, are \mbox{$\alpha_{J2000}$ = $17^h$47$^m22^s$} and $\delta_{J2000} = -28^{\circ}21^{'}27^{''}$. The observations were performed using the position switching mode, with the OFF position located at ($-885''$, $+290''$) from G\,+0.693. The line intensity of the spectra is given in T$_A^*$ as the molecular emission toward G\,+0.693 is extended over the beam of the telescope. Here, we briefly describe the observations. For more details, we refer to \citet{zen2020}, \citet{rod2021a} and \citet{riv2022b}.

For the GBT observations, the Ku-band receiver was connected to the spectrometer, providing four 200 MHz spectral windows with a spectral resolution of 195 kHz, corresponding to a velocity resolution of 2.2$-$8.6 km s$^{-1}$, and covering a frequency range between 12 GHz and 26 GHz. 

For the Yebes 40m observations, we used the Q-band (7 mm) HEMT receiver developed within the Nanocosmos project \citep{ter2021}. The receiver was connected to 16 fast Fourier transform spectrometers (FFTSs), providing a channel width of 38 kHz and a bandwidth of 18.5 GHz per polarisation, covering the frequency range between 31.3 GHz and 50.6 GHz. The spectra were smoothed to a resolution of 251 kHz, equivalent to a velocity resolution of 1.5$-$2.4 km s$^{-1}$. 

The IRAM 30m data used in this work correspond to observations made in the 1mm, 2mm, and 3mm bands and cover several frequency ranges from 71.8$-$116.7 GHz, 124.8$-$175.5 GHz, and 199.8$-$238.3 GHz. The EMIR receivers were connected to the FFTS, providing a spectral resolution of 200 kHz. The spectra were smoothed to velocity resolutions of 1.0$-$2.6 km s$^{-1}$, depending on the frequency.

 \begin{table*}
 \centering    
 \resizebox{1.2\columnwidth}{!}{     
  \begin{threeparttable}  
\caption{Covered rotational transitions of SiC$_2$, SiS, and Si$^{18}$O.}              
\label{table:observed_lines}                        
\begin{tabular}{lccccl}          
\hline \hline
\multicolumn{1}{l}{Molecule} & \multicolumn{1}{c}{Transition}  &\multicolumn{1}{c}{Frequency} & \multicolumn{1}{c}{log\,$I$ (300 K)} & \multicolumn{1}{l}{$E_u$} & \multicolumn{1}{l}{Blending}\\
&             & \multicolumn{1}{c}{(GHz)} & \multicolumn{1}{c}{(nm$^{2}$ MHz)} & \multicolumn{1}{c}{(K)} &  \\
\hline                                 
SiC$_2$ & $2_{0,2} - 1_{0,1}$ & 47.06482  & $-$4.49&  3.4 & unblended \\
& $4_{0,4} - 3_{0,3}$ &93.06363 &       $-$3.61 & 11.2 &  unblended \\
& $4_{2,2} - 3_{2,1}$ &95.57938 &       $-$3.72 & 19.2 & unblended\\
& $5_{0,5} - 4_{0,4}$ &  115.38236 & $-$3.33 & 16.7 & blended with NCCNH$^+$ \\
& $6_{0,6} - 5_{0,5}$ & 137.18078 & $-$3.11 & 23.3  & blended with $c$-HCOOH\\
& $6_{2,4} - 5_{2,3}$ & 145.32587 &  $-$3.12 & 31.9 & unblended \\ 
SiS & $1-0$ & 18.15488 & $-5.02$ & 0.9 & unblended \\
 & $2-1$& 36.30963 & $-$4.12 & 2.6 & unblended\\
 & $4-3$ & 72.61810 & $-3.22$ & 8.7 & unblended\\
 & $5-4$ & 90.77155 & $-2.93$ & 13.0 & unblended \\
 & $6-5$ & 108.92427 & $-2.70$ & 18.3& unblended \\
 & $7-6$ & 127.07614 & $-2.51$ &24.4 &unblended \\
 & $8-7$ & 145.22699 & $-2.35$ & 31.4& unblended\\
 & $9-8$ & 163.37670 & $        -2.20$ &        39.2&unblended \\
Si$^{18}$O &$1-0$ & 40.35276 &  $-$3.47&  1.9 &  unblended \\
& $2-1$ & 80.70493 &  $-$2.57 & 5.8 & unblended \\
& $4-3$ & 161.40488 &  $-$1.69 &  19.3 & blended with CH$_3$OCHO \\
& $5-4$ & 201.75148 & $-$1.41& 20.0 &  unblended\\
\hline                                           
\end{tabular}
\begin{tablenotes}
      \footnotesize
      \item Note: For the spectroscopic information, see CDMS for SiC$_2$ and Si$^{18}$O and JPL for SiS.
    \end{tablenotes} 
    \end{threeparttable}
    }
\end{table*}

\section{Analysis and results} \label{section:results}

\subsection{LTE analysis}\label{section:lte}

To identify and analyze the molecular lines within our spectral survey of G\,+0.693, we used the Spectral Line Identification and Modelling (SLIM) package within MADCUBA\footnote{Madrid Data Cube Analysis (MADCUBA) is software developed in the Centre of Astrobiology (CAB) to analyze astronomical datacubes and spectra.} \citep{mar2019}, which includes the spectroscopic information of the Cologne Database for Molecular Spectroscopy (CDMS, \citealt{mul2005}) and the Jet Propulsion Laboratory (JPL, \citealt{pic1998}) catalogues for different molecular species. 

SLIM generates synthetic spectra assuming LTE conditions and uses an algorithm (AUTOFIT) that provides the best nonlinear least-squares fit to the data. By performing the fit, we aim to derive the physical parameters of the molecular emission, namely the molecular column density ($N$), the excitation temperature ($T_{\rm ex}$), the velocity with respect to the Local Standard of Rest ($v_{\rm LSR}$), and the line width (full width at half maximum, FWHM). To evaluate whether or not the molecular transitions are contaminated with emission from other species, we consider the predicted LTE emission from more than 120 molecules already identified toward G\,+0.693. 

\begin{figure}
\centering
    \includegraphics[scale=0.5]{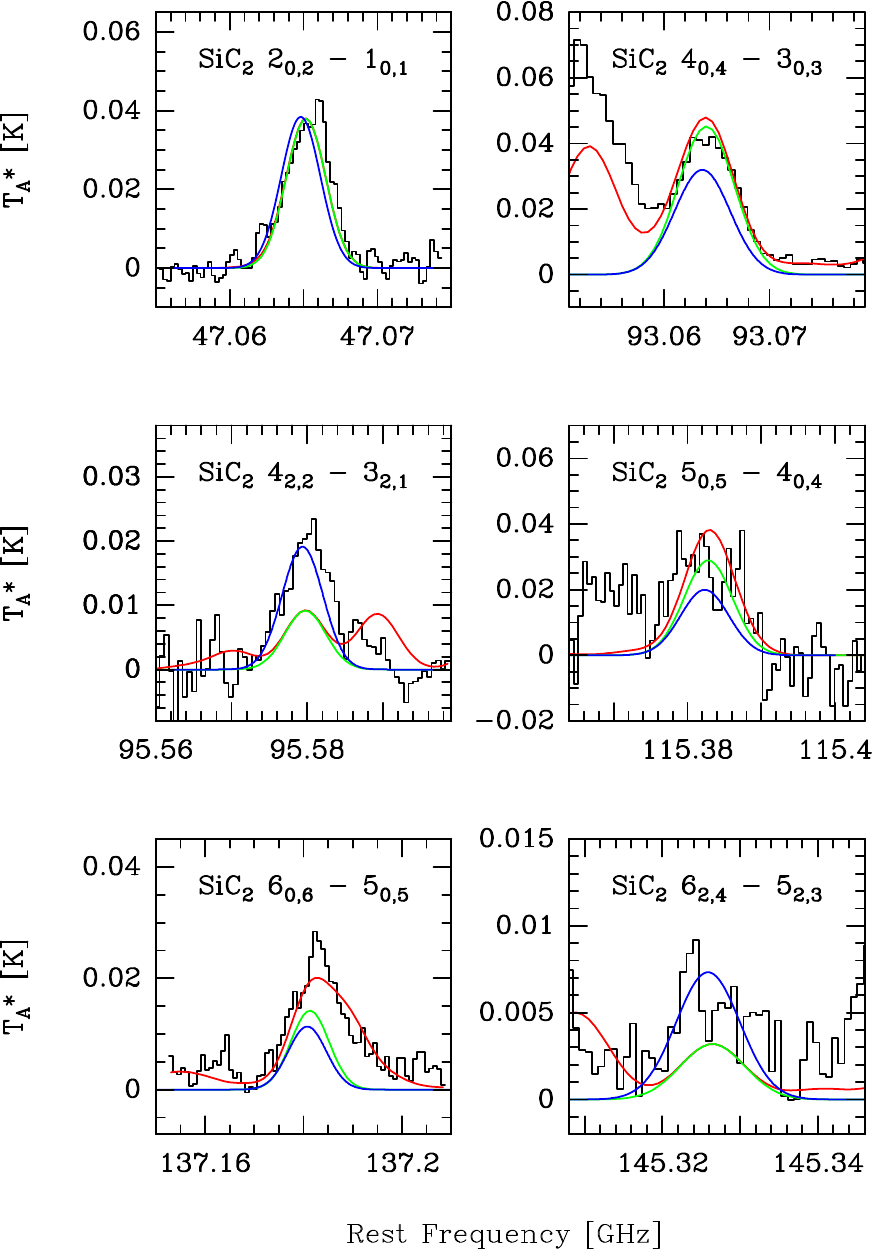}
    \caption{Observed spectra of SiC$_2$ toward G\,+0.693 (black histogram). The green lines correspond to the synthetic spectra calculated assuming LTE. The red lines show the synthetic spectra that account for the emission of all the molecules identified toward G\,+0.693, including SiC$_2$. The blue lines correspond to the line profiles calculated using \texttt{Radex}.}
    \label{fig:sic2}
\end{figure}

We detect SiC$_2$ toward G\,+0.693 through six rotational transitions (Table~\ref{table:observed_lines}). This is the first time this molecule has been reported in the ISM. The observed lines are shown in Fig.~\ref{fig:sic2}. The $5_{0,5} - 4_{0,4}$ transition of SiC$_2$ at $\sim$115 GHz is slightly blended with NCCNH$^+$, and the $6_{0,6} - 5_{0,5}$ transition at $\sim$137 GHz is blended with $c$-HCOOH. An examination of the number of identified and unidentified lines within a velocity range of $\pm$200 km s$^{-1}$ around the detected SiC$_2$ transitions reveals that about 15\% of the lines observed in those frequency ranges are unidentified.

To carry out the LTE fit, we fixed the value of the FWHM to 20 km s$^{-1}$ for the AUTOFIT algorithm to converge. This value is typical of molecular emission in the region (e.g., \citealt{zen2018,riv2022c}) and reproduces the line profiles of SiC$_2$ extremely well. The derived physical parameters together with their corresponding uncertainties are listed in Table~\ref{table:parameters}. For the parameters that were fixed, no uncertainties are associated with the derived values. The best LTE fit is obtained for $N=(1.02\pm0.04)\times10^{13}$ cm$^{-2}$, $T_{\rm ex} = 5.9\pm0.2$ K and $v_{\rm LSR} = 67.8\pm 0.4$ km s$^{-1}$. 

From Fig.~\ref{fig:sic2}, it is clear that the LTE fit underestimates the $4_{2,2}-3_{2,1}$ line at $\sim$ 95 GHz and the $6_{2,4}-5_{2,3}$ line at $\sim$ 145 GHz. We note that a higher $T_{\rm ex}$ would not explain this excess of emission because it would overpredict the $6_{0,6} - 5_{0,5}$ line at $\sim$137 GHz, which is well fit. As analyzed in Sect.~\ref{section:non-lte}, this is due to a nonLTE excitation effect of the SiC$_2$ line emission as expected due to the rather low H$_2$ density found in this source. 

\begin{figure}
\centering
    \includegraphics[scale=0.5]{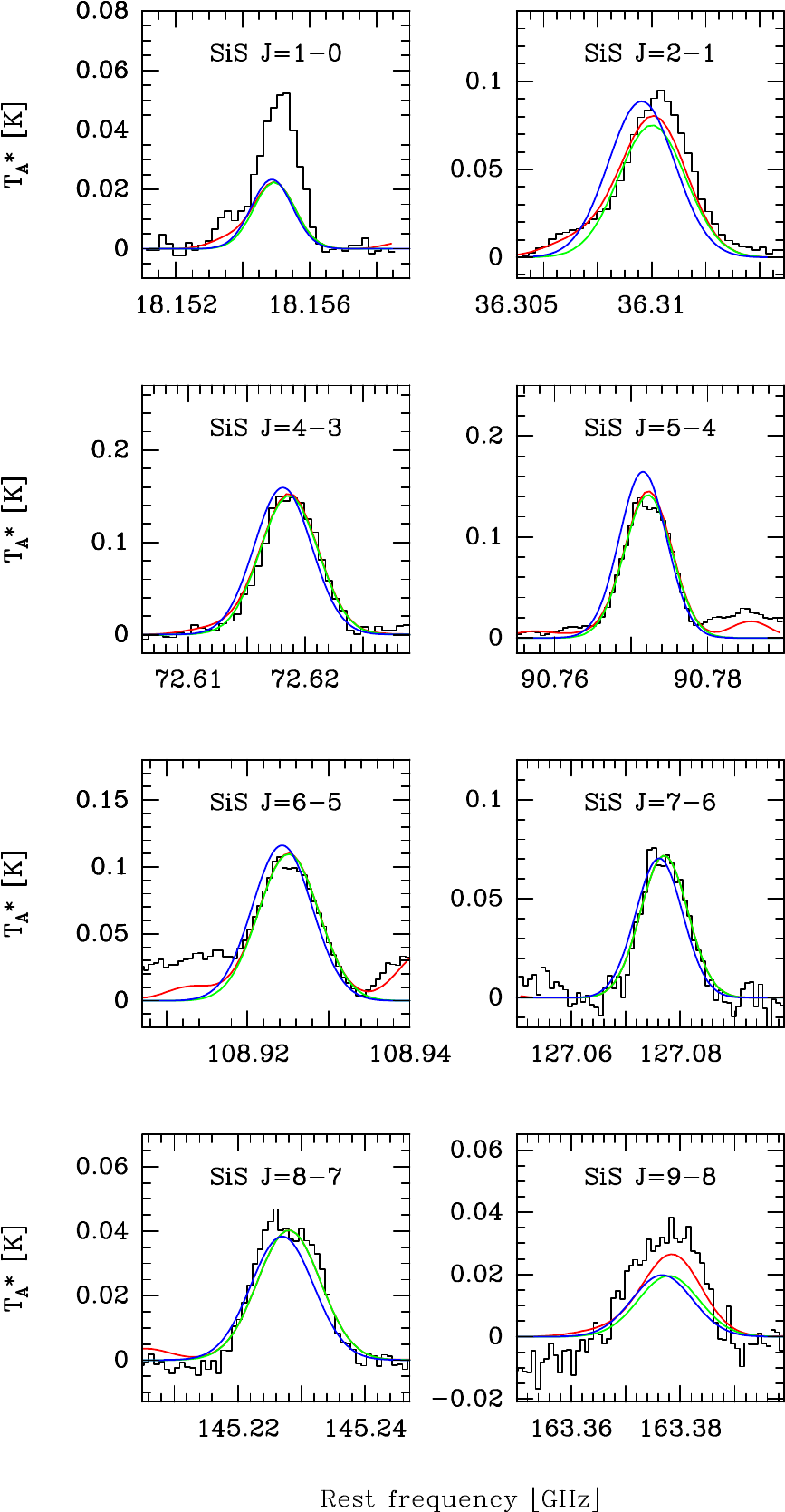}
    \caption{Observed spectra of SiS toward G\,+0.693 (black histogram). The green lines correspond to the synthetic spectra calculated assuming LTE. The red lines show the synthetic spectra that account for the emission of all the molecules identified toward G\,+0.693. The blue lines correspond to the line profiles calculated using \texttt{Radex}.}
    \label{fig:sis}
\end{figure}

 We also searched for other Si-bearing molecules toward G\,+0.693, specifically SiS and SiO. We detect eight rotational transitions of SiS (Table~\ref{table:observed_lines}), which are shown in Fig.~\ref{fig:sis}. All of the transitions of SiS are unblended. The derived parameters of the best LTE fit are \mbox{$N=(5.3\pm0.1)\times10^{13}$ cm$^{-2}$}, $T_{\rm ex} = 8.0\pm0.1$ K, $v_{\rm LSR} =66.8\pm 0.2$ km s$^{-1}$ and FWHM = $24.0\pm0.3$ km s$^{-1}$ (see Table~\ref{table:parameters}). As in the case of SiC$_2$, in Sect.~\ref{section:non-lte} we investigate whether the emission from this molecule suffers from nonLTE excitation effects.
 
\begin{figure}
\centering
    \includegraphics[scale=0.5]{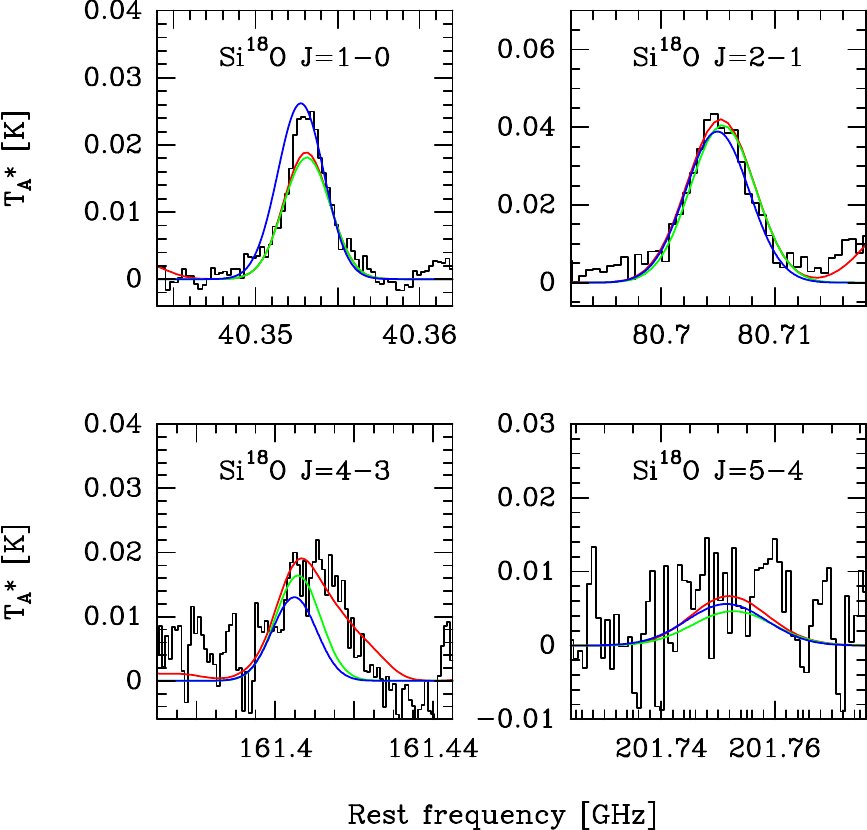}
    \caption{Observed spectra of the isotopolog Si$^{18}$O toward G\,+0.693 (black histogram). The green lines correspond to the synthetic spectra calculated assuming LTE. The red lines show the synthetic spectra that account for the emission of all the molecules identified toward G\,+0.693. The blue lines correspond to the line profiles calculated using \texttt{Radex}.}
    \label{fig:sio}
\end{figure}

The SiO emission line is optically thick in the GC, and therefore we focus on the optically thin isotopolog, Si$^{18}$O. We detect four rotational transitions of Si$^{18}$O (Table~\ref{table:observed_lines}), which are shown in Fig.~\ref{fig:sio}. All of the transitions are clean apart from the $J=4-3$ transition at $\sim$161 GHz, which is slightly blended with CH$_3$OCHO. We derive best-LTE-fit parameters of \mbox{$N=(2.9\pm0.1)\times10^{12}$ cm$^{-2}$}, \mbox{$T_{\rm ex}=5.5\pm0.2$ K},  $v_{\rm LSR}=67.3\pm 0.4$ km s$^{-1}$, and FWHM = $24.8\pm1.1$ km s$^{-1}$ (Table~\ref{table:parameters}). From this, we estimate a SiO column density of \mbox{$N=(7.2\pm0.3)\times10^{14}$ cm$^{-2}$} using the measured isotopic ratio for oxygen in the GC of $^{16}$O/$^{18}$O = 250 \citep{wil1999}.
  
 The SiO column density can also be estimated from the $^{29}$SiO and $^{30}$SiO isotopologs and we therefore analyzed their transitions. Their derived parameters are found in Table~\ref{table:parameters}. Assuming isotopic ratios for silicon $^{29}$Si/$^{28}$Si = 0.095 \citep{wol1980} and $^{29}$Si/$^{30}$Si = 1.5 \citep{wil1994} which results in $^{30}$Si/$^{28}$Si = 0.063, we obtain SiO column densities of  $N=2.6\times10^{14}$ cm$^{-2}$ using $^{29}$SiO, and $N=2.8\times10^{14}$ cm$^{-2}$ using $^{30}$SiO. These values are lower than the one obtained using Si$^{18}$O ($N=7.2\times10^{14}$ cm$^{-2}$; see Table~\ref{table:parameters}), which suggests that the emission of these isotopologs is moderately optically thick. Therefore, we used Si$^{18}$O to perform the analysis of SiO. No collisional rate coefficients are available for Si$^{18}$O to carry out the nonLTE analysis of this emission. However, as presented in Sect.~\ref{section:non-lte}, we used the collisional data of SiO as a good approximation. 
 
 The values derived for $T_{\rm ex}$, $v_{\rm LSR}$, and FWHM for all these species are in agreement with those obtained previously for other molecular species toward this cloud ($T_{\rm ex}\sim\,5-20$ K, $v_{\rm LSR}$ $\sim$ 68 km s$^{-1}$, and line widths $\sim 20$ km s$^{-1}$, e.g., \citealt{req2006, zen2018, riv2020, rod2021a}). Additionally, we searched for the emission of SiC, Si$_2$C, and SiN; however, we do not detect these molecules toward G\,+0.693. The derived upper limits on the column densities are presented in Table~\ref{table:non-detections}.

\subsection{NonLTE analysis}\label{section:non-lte}
As mentioned in Sect.~\ref{section:lte}, the $4_{2,2}-3_{2,1}$ and the $6_{2,4}-5_{2,3}$ lines of SiC$_2$ are underestimated by the LTE model of MADCUBA. The kinetic temperature of the gas toward the G+0.693 cloud is $T_{\rm kin}$$\sim$70$-$150 K, and the H$_2$ gas density is a few 10$^4$ cm$^{-3}$ as determined by \citet{zen2020} using emission of Class I methanol masers. We note that the maser emission is extended across the whole Sgr B2 molecular cloud (e.g., \citealt{lie1996}), which indicates that the physical properties of the gas responsible for the methanol maser emission in G\,+0.693 are similar to the ones found across the whole Sgr B2 cloud. In addition, the H$_2$ volume gas densities have also been measured toward other giant molecular clouds (GMCs) in the GC and are also a few 10$^{-4}$ cm$^{-3}$, as reported by \citet{gue1983} and \citet{gus2004}. Given that the H$_2$ volume density of the gas in G\,+0.693 is low, the excitation of heavy molecules such as SiC$_2$ is expected to be nonLTE. We therefore used the online nonLTE radiative transfer code \mbox{\texttt{Radex} \citep{van2007}}, which allows us to check for nonLTE excitation and to constrain the physical properties of the molecular gas.

To run \texttt{Radex}, we set the cosmic microwave background radiation temperature to 2.73 K, the H$_2$ gas density to a typical value of $2\times10^{4}$ cm$^{-3}$ for the GC (\citealt{rod2000,gus2004,zen2020}), and the column density, $N$, similar to the value provided by MADCUBA, that is $N = 1.2\times10^{13}$ cm$^{-2}$ for SiC$_2$ and $N = 5.0\times10^{13}$ cm$^{-2}$ for SiS. In the case of Si$^{18}$O, the collisional rate coefficients are not available; nevertheless, we performed the nonLTE calculations using the collisional data of SiO, because they are expected to be similar to those of Si$^{18}$O. We input $N = 2.0\times10^{12}$ cm$^{-2}$ for Si$^{18}$O, and find that a higher H$_2$ gas density of $6.0\times10^{4}$ cm$^{-3}$ is needed for this molecule. We then varied $T_{\rm kin}$ and compared the line intensities observed with those predicted by \texttt{Radex}. The results are overplotted in blue in Fig.~\ref{fig:sic2} for SiC$_2$, Fig.~\ref{fig:sis} for SiS, and Fig.~\ref{fig:sio} for Si$^{18}$O.  

From the calculations, we obtain a T$_{\rm kin}$ = 150 K for SiC$_2$ and Si$^{18}$O, while we infer a lower T$_{\rm kin}$ = 90 K for SiS. The derived T$_{\rm kin}$ are similar to values previously reported for CH$_3$CN (70-150 K, \citealt{zen2018,zen2020}) toward the same source. We also note that the derived T$_{\rm ex}$ is well below T$_{\rm kin}$, which indicates that the molecular emission in G\,+0.693 is subthermally excited, as was found for other species in this region owing to the low H$_2$ gas density \citep{req2006, req2008, zen2018}.

 Indeed, the $4_{2,2}-3_{2,1}$ and $6_{2,4}-5_{2,3}$ lines of SiC$_2$ are underestimated by the LTE model of MADCUBA; however, they are well explained when the nonLTE excitation is considered (see Fig.~\ref{fig:sic2}). This suggests that the transitions substantially deviate from quasi-thermal excitation, where quasi-thermal refers to subthermal excitation where the molecular emission can be well described using a single $T_{\rm ex}$ (see \citealt{gol1999} for the definition of quasi-thermal excitation).
 
On the other hand, for SiS, there is a good agreement between the line intensities calculated by \texttt{Radex} and those predicted by MADCUBA, which points to quasi-thermal excitation similar to that observed in other molecules in this source (see Fig.~\ref{fig:sis}). We notice that for the $J=1-0$ transition of SiS, there seems to be a discrepancy between the observed and the predicted line intensities. This discrepancy could be due to population inversion and suprathermal excitation of the SiS $J=1-0$ transition. We note that these effects are expected to be found in the low$-J$ transitions of linear molecules, such as SiS and even CO, which are a consequence of collisional propensity rules \citep{koe1980}. Indeed, our \texttt{Radex} calculations show that the T$_{\rm ex}$ and the calculated optical depth, $\tau$, obtained for the SiS $J=1-0$ line are negative, which suggests that suprathermal excitation may be at play.
It is also possible that the mismatch between the observed and predicted SiS $J=1-0$ lines is due to the $J=1-0$ transition being blended with emission of a molecular species that is not yet identified. We remark that this mismatch cannot be related to a beam effect, because the beam sizes of the GBT and Yebes 40m telescopes at the corresponding frequencies are comparable. 
 
For Si$^{18}$O, the line intensities calculated by \texttt{Radex} are similar to those predicted by MADCUBA, which indicates a quasi-thermal excitation for this molecule (see Fig.~\ref{fig:sio}). We notice that the Si$^{18}$O $J=1-0$ transition is slightly better reproduced when considering the nonLTE approach, which suggests that nonLTE effects play a role in the excitation of this transition.
  
Finally, we note that the excitation temperatures calculated in Sect.~\ref{section:lte} are lower than the gas kinetic temperatures derived here (T$_{\rm ex}$ < T$_{\rm kin}$) because the emission is subthermally excited due to the low H$_2$ gas densities in the cloud. However, as was found for other molecules in G+0.693, the observed lines are well reproduced when assuming LTE and a single T$_{\rm exc}$, although a better match is achieved for some species such as SiC$_2$ when a nonLTE excitation analysis is carried out.

\subsection{Fractional abundances relative to H$_2$}
  
We estimate the fractional abundances in G\,+0.693, $X$, that is, the abundance of the molecules relative to H$_2$, assuming a column density of \mbox{N$_{\rm H_2}$ = $1.35\times10^{23}$ cm$^{-2}$} as measured toward G\,+0.693 \citep{mar2008}. The abundances are reported in Table~\ref{table:parameters}.  We derive \mbox{$X_{\rm SiC_{2}}=7.5\times10^{-11}$}, \mbox{$X_{\rm SiS}$ = 3.9 $\times 10^{-10}$}, and \mbox{$X_{\rm Si^{18}O}$ = $2.1\times10^{-11}$}. We also estimate \mbox{$X_{\rm SiO}$ = $5.3\times10^{-9}$}.  The SiO and SiS abundances of a few $\sim$10$^{-9}$ and $\sim$10$^{-10}$, respectively, with SiO being more abundant, are consistent with findings of previous observational studies in the GC \citep{dic1981, mar1997, hut1998, rod2006, amo2009, min2015}, while SiC$_2$ appears to be the least abundant in G\,+0.693 as compared to the other Si-bearing species. We also calculated the abundance ratios of SiO with respect to SiS and SiC$_2$, and find that SiO/SiS $\sim$14 and SiO/SiC$_2 \sim70$ toward G\,+0.693. 

For the nondetected silicon-bearing species (SiC, Si$_2$C and SiN), we calculate upper limits to the fractional abundances using the rms noise level of the brightest and least contaminated transition. The upper limits, as well as the transition used as a reference, are presented in Table~\ref{table:non-detections}. We obtain \mbox{X$_{\rm SiC}$ $\le 5.8\times 10^{-12}$} and \mbox{SiO/SiC $\geq905$}, \mbox{X$_{\rm Si_{2}C}$ $\le4.7 \times 10^{-11}$} and \mbox{SiO/Si$_2$C $\geq115$}, and finally \mbox{X$_{\rm SiN}$ $\le9.6\times 10^{-11}$} and SiO/SiN $\geq55$.

 \begin{table*}
 \centering    
 \resizebox{1.3\columnwidth}{!}{     
  \begin{threeparttable}  
\caption{Derived parameters of SiC$_2$, SiS, $^{29}$SiO, $^{30}$SiO, and Si$^{18}$O.}           
\label{table:parameters}                        
\begin{tabular}{lcccccc}          
\hline\hline                     
Molecule & T$_{\rm ex}$ & v$_{\rm LSR}$ & FWHM & N & $X$  &    $X_{SiO}/X$   \\    
 & (K) &  (km s$^{-1}$)  & (km s$^{-1}$) & (cm$^{-2}$)  & & \\
\hline                                   
SiC$_2$  & $5.9\pm0.2$ & $67.8\pm0.4$ & 20.0$^{a}$ &  (1.02 $\pm$ 0.04) $\times\,10^{13}$   & 7.5 $\times10^{-11}$ & 70\\
SiS  & $8.0\pm0.1$& $66.8\pm0.2$ & $24.0\pm0.3$ &  (5.3 $\pm$ 0.1) $\times\,10^{13}$ & 3.9 $\times 10^{-10}$ & 14\\
$^{29}$SiO &  $4.6\pm0.2$ & $68.8\pm0.4$ & $23.2 \pm 0.4$ & ($2.5\pm0.1$) $\times\,10^{13}$ & 1.9 $\times 10^{-10}$ & 28\\
$^{30}$SiO  & $4.6\pm0.1$ & $69.1\pm0.2$ &  $23.1\pm 0.5$ &  ($1.8\pm0.1$) $\times\,10^{13}$  &  1.3 $\times 10^{-10}$ & 41 \\
Si$^{18}$O  & $5.5\pm0.2$ & $67.3\pm0.4$ & $24.8\pm1.1$ & (2.9 $\pm$ 0.1)$\,\times$\,10$^{12}$ & $2.1\times10^{-11}$ &250 \\
SiO$^{b}$ & $-$ & $-$ & $-$ & (7.2 $\pm$ 0.3)$\,\times$\,10$^{14}$ & $5.3\times10^{-9}$ & 1\\
\hline                                           
\end{tabular}
\begin{tablenotes}
      \small
      \item $^{a}$ Fixed value.
      \item $^{b}$ Values obtained assuming an isotopic ratio $^{16}$O/$^{18}$O = 250 \citep{wil1999}.
    \end{tablenotes} 
    \end{threeparttable}
    }
\end{table*}

\begin{table*}
\centering    
\resizebox{1.3\columnwidth}{!}{
  \begin{threeparttable}  
\caption{Derived upper limits of SiC, Si$_2$C, and SiN.}              
\label{table:non-detections}    
\centering                                   
\begin{tabular}{lcccccc}          
\hline\hline                       
Molecule & Transition & Frequency &  rms              & N                       & $X$ & $X_{SiO}/X$    \\    
               &                  & GHz & (mK) &  (cm$^{-2}$)  & \\
\hline 
SiC & 3$_{1,4} - 2_{-1,3}$ &  157494.10 & 8 & $\le7.9\times 10^{11}$ & $\le 5.8\times 10^{-12}$ & $\geq905$\\                                  
Si$_2$C & 5$_{1,5} - 4_{0,4}$ & 100120.66 &4    & $\le6.3\times 10^{12}$   & $\le4.7 \times 10^{-11}$ &  $\geq115$\\
SiN         &   1$_{1,1} - 0_{1,2}$  & 43098.66    &   2        &  $\le1.3\times 10^{13}$       & $\le9.6\times 10^{-11}$ & $\geq55$\\
\hline                                           
\end{tabular}
\begin{tablenotes}
      \small
      \item 
    \end{tablenotes}
    \end{threeparttable}
    }
\end{table*}

\section{Discussion}\label{section:discussion}

\subsection{Comparison of abundances with those measured in other environments}

\begin{figure}
\centering
    \includegraphics[width=\columnwidth]{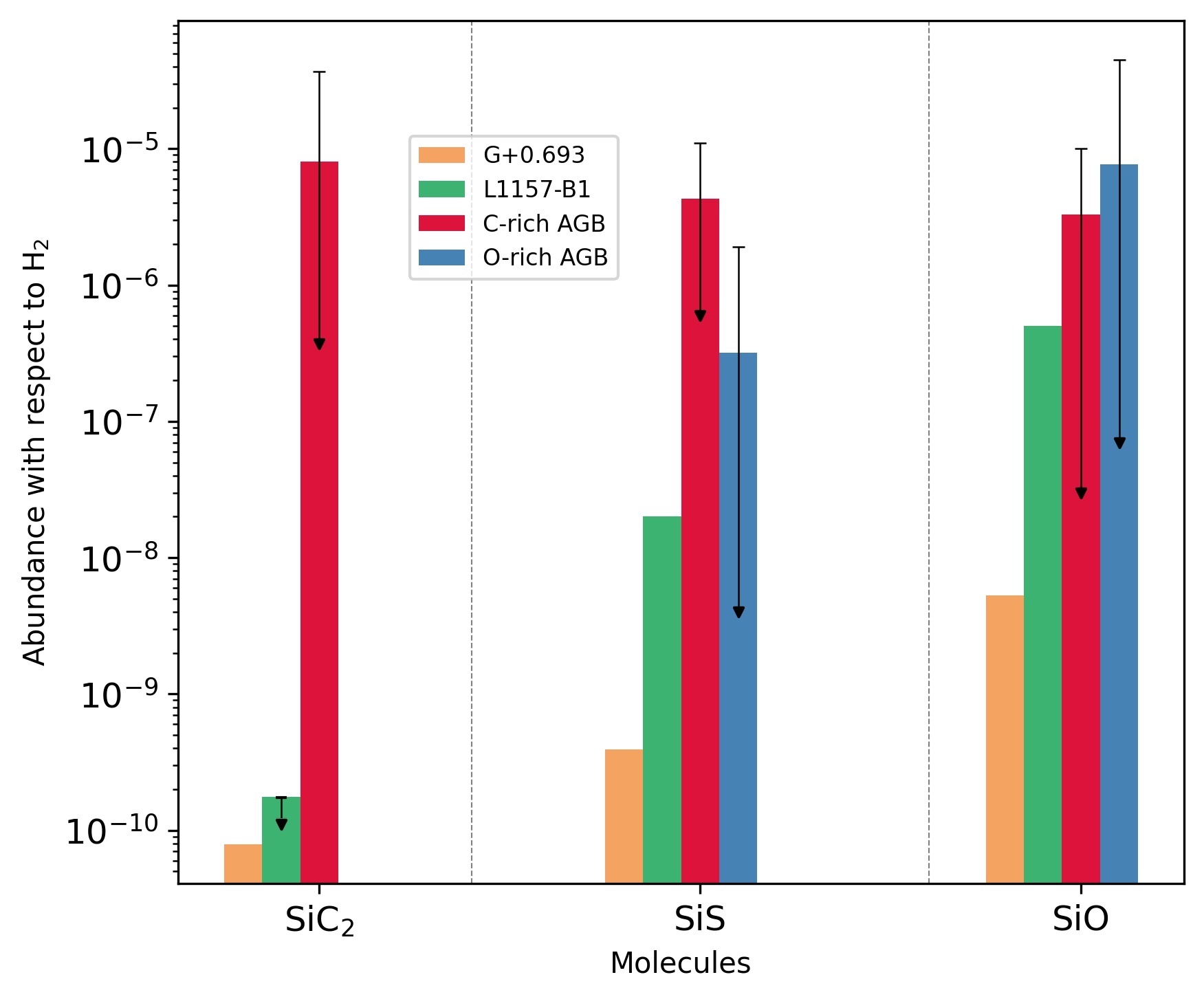}
    \caption{Bar plot comparing the fractional abundance of SiC$_2$, SiS, and SiO relative to H$_2$ in different environments. For G\,+0.693, the abundances are obtained from this work. For L1157-B1, the SiO and SiS abundances are taken from \citet{pod2017}, while the upper limit on the SiC$_2$ abundance is derived in this work using the ASAI survey \citep{asai2018}. The abundances of the carbon-rich AGB stars and oxygen-rich AGB stars are from \citet{mas2018,mas2019} and \citet{mas2020}, respectively. For the AGB stars, the height of the bar corresponds to the average fractional abundance derived in the studied samples, while the error bar refers to the wide range of values found. Downward-pointing arrows represent upper limits to the abundance.}
    \label{fig:abundance_comparison}
\end{figure}

To evaluate how efficiently the Si-bearing molecules are produced in G\,+0.693, we compare  their abundances in Fig.~\ref{fig:abundance_comparison} with those derived in other chemically rich environments, namely the protostellar shock L1157-B1, and carbon- and oxygen-rich AGB stars. The comparison with AGB stars is carried out to elucidate whether interstellar SiC$_2$ is a product of circumstellar envelopes of evolved stars, where dust grains act as a reservoir carrying the molecule all the way into the ISM, until released directly into the gas phase through sputtering.

The derived fractional abundances in G\,+0.693 are plotted in orange. In green, we show the SiO and SiS abundances toward the protostellar shock L1157-B1 calculated by \citet{pod2017}. We also searched for SiC$_2$ in L1157-B1 using the spectral line survey of ASAI (Astrochemical Surveys at IRAM; \citealt{asai2018}); however, we find no detection of SiC$_2$ emission toward this source down to an rms noise level of 3 mK; we therefore derived an upper limit on the SiC$_2$ abundance of <$1.8\times10^{-10}$ assuming \mbox{N$_{\rm H_2}$= $9\times10^{20}$ cm$^{-2}$} \citep{pod2017}, which is also plotted in Fig.~\ref{fig:abundance_comparison} in green. We show the average fractional abundances of SiC$_2$, SiS, and SiO derived for 25 C-rich AGB stars studied by \citet{mas2018,mas2019}  in red and the average fractional abundances of SiS and SiO derived in 30 O-rich envelopes by \citet{mas2020}  in blue.

From Fig.~\ref{fig:abundance_comparison}, we find that the fractional abundances of the Si-bearing molecules in G\,+0.693 are the lowest compared to other environments. This may indicate that their formation is less efficient toward this cloud, while the highest abundances are found toward evolved stars. This is not surprising as these molecules form in the densest ($\sim$10$^{13}-$10$^{15}$ cm$^{-3}$) and hottest ($\sim$ 2000 $-$ 3000 K) parts of the inner envelopes of AGB stars under LTE conditions, and together are thought to lock up a significant fraction of the silicon elemental abundance \citep{mas2018,mas2019}. 

\begin{figure*}
\centering
\includegraphics[width=0.4\textwidth]{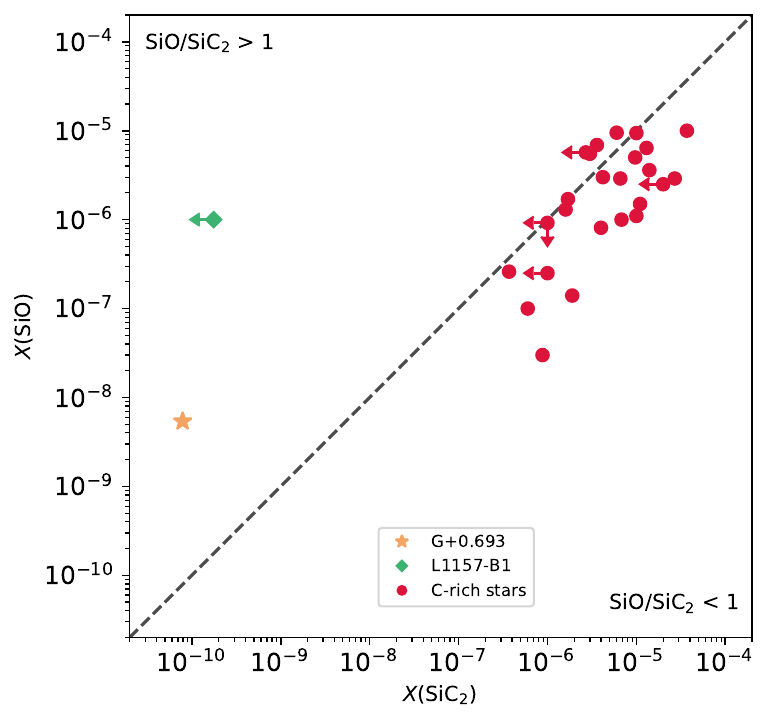} \includegraphics[width=0.4\textwidth]{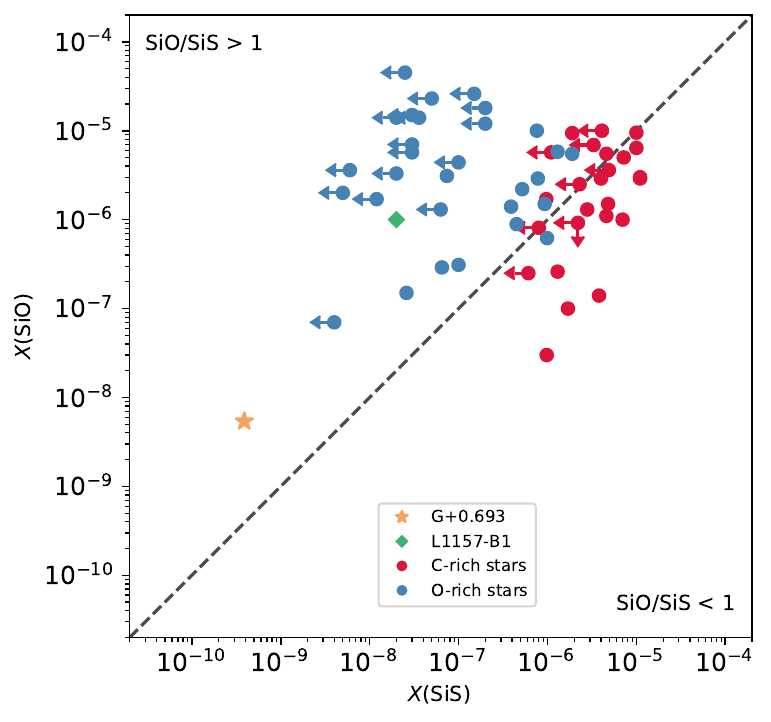}
\caption{Comparison of abundances between different pairs of molecules. The plots show the derived fractional abundances relative to H$_2$ of SiO vs. SiC$_2$ \textit{(left panel)}, and SiO vs. SiS \textit{(right panel)}. The sources with nondetections are denoted with arrows. The dashed line represents equal abundances of molecules.}
 \label{fig:abundance_ratio}
\end{figure*}

In Fig.~\ref{fig:abundance_ratio}, we plot the fractional abundance of SiO against that of SiC$_2$ (left panel) and the fractional abundance of SiO against that of SiS (right panel), in the different environments. We find that for G\,+0.693 and \mbox{L1157-B1}, SiO is more abundant than SiC$_2$ as indicated by the fact that the sources lie in the \mbox{SiO/SiC$_2$ > 1} region. This behavior is different than what is found for the envelopes of carbon evolved stars where, for most of the studied sample, SiC$_2$ tends to be more abundant than SiO. This suggests that in the ISM, SiO is likely to lock up silicon more efficiently than SiC$_2$. This is probably due to the fact that SiO is more refractory than SiC$_2$ \citep{lod2003}, better surviving the journey from the ejected circumstellar material to the ISM. In the right panel of Fig.~\ref{fig:abundance_ratio}, the comparison between SiO and SiS shows that the data points of G\,+0.693 and \mbox{L1157-B1} lie in the \mbox{SiO/SiS > 1} side. This is somewhat closer to the silicon chemistry of oxygen-rich envelopes, where SiO tends to be predominantly more abundant than SiS, although in the ISM it is clearly of lower order, implying that lower amounts of silicon are carried there. 
 
\subsection{Exploring the origin of Si-bearing molecules in the ISM}

The presence of Si-bearing molecules in interstellar clouds is attributed to violent events that can shatter the dust grains through ion sputtering or grain--grain collisions, effectively releasing the Si-containing species either directly from dust grains to the gas phase, or releasing Si atoms that can then immediately react to form molecules. Silicon monoxide is widely observed in outflows affected by shocks and is considered to be a powerful shock tracer \citep{mar1992}. Because of this link, its formation is believed to be due to shocks that release SiO molecules from the grains directly into the gas phase (e.g., \citealt{jim2005}) or by releasing Si atoms that react through \mbox{Si + O$_2$ → SiO + O} and \mbox{Si + OH → SiO + H} \citep{lan1990,sch1997}. Observational studies in circumstellar envelopes of AGB stars found an SiO abundance depletion with increasing envelope density, suggesting efficient incorporation of this molecule onto dust grains (e.g., \citealt{sch2006,mas2018,mas2020}). It therefore appears that SiO is a product of the release from dust grains in shocked regions.

On the other hand,  SiS emission has been observed less in shocked regions and its origin in the ISM is not adequately understood. SiS emission has only been found in a few star-forming regions where outflows are present, such as Sgr B2 and Orion KL, and more recently in L1157-B1, an outflow driven by a low-mass protostar (\citealt{dic1981,ziu1988,ziu1991,ter2011,pod2017}). In their observational study of Si-bearing molecules in L1157-B1, \citet{pod2017} found a strong gradient across the shock, where SiS is only detected at the head of the outflow cavity and not detected at the shock impact region, implying that SiS is not directly released from the grains but is instead formed through gas-phase processes after part of the atomic silicon is ejected. It is worth noting that, as opposed to the case for SiO, observational studies of evolved stars did not find a strong trend of decreasing SiS abundance with increasing envelope density, which would be expected if adsorption onto dust grains were important for this molecule (\citealt{sch2007,mas2019,mas2020}). \citet{ros2018} suggested the formation of SiS in the gas phase through the reactions \mbox{SiH + S and SiH + S$_2$}. \citet{zan2018} proposed another path that involves the reaction of atomic silicon with either SO or SO$_2$. More recently, \citet{man2022} investigated the reaction \mbox{S$^{+}$($^{4}$S) + SiH$_2$($^1$A$_1$)} as a possible route as well. 

Contrary to SiO and SiS, detections of SiN are rare, where apart from evolved stars \citep{tur1992,bor2022}, its emission has been observed only toward Sgr B2 (M) in absorption \citep{sch2003}. Our current understanding of its formation in the ISM is limited. However, if enough Si is available, then one of its formation pathways could be \mbox{Si + NH $\rightarrow$ SiN + H} (\citealt{rov1988,sch2003}). 

Efforts have also been made to detect SiC in the ISM \citep{sch1997, ter2011}. We do not detect this molecule toward G\,+0.693 either. \citet{roc2022} performed a theoretical calculation of the rate constants for the gas-phase reaction C$_2$ + Si → SiC + C for temperatures between 2000 and 5000 K, and found that the efficiency of the reaction has a positive dependence on the temperature, where it increases steeply beyond 3000 K. This indicates that SiC is unlikely to form in the ISM through this route. In evolved stars, SiC was detected in IRC\,+10216 by \citet{cer1989} with line profiles indicating that it is formed in the outer layers of the envelope, probably as a photodissociation product of SiC$_2$.

\subsection{Formation of SiC$_2$ in the ISM}
Previous to this work, SiC$_2$ has not been detected in shocked regions in the ISM. The reason for this could be due to the lines of SiC$_2$ not targeted by observations as often as other Si-bearing species, such as SiO and SiS. Another explanation could simply be that SiC$_2$ is less abundant. The detection of SiC$_2$ toward G\,+0.693 in this work leads us to speculate about two possible scenarios for the presence of this molecule in the ISM. Being a likely precursor of SiC dust \citep{mas2018}, SiC$_2$ could be locked onto dust grains and then released directly into the gas phase as a consequence of sputtering caused by shocks in the region. Low- and moderate-velocity shocks can liberate molecules from grains and into the gas phase without destroying the molecular bonds \citep{req2006}. However, we note that solid SiC dust has not yet been detected in the ISM \citep{che2022}.

It is worth exploring the efficiency of the formation of interstellar SiC$_2$ in the gas phase under the typical physical conditions of the GC. The UMIST Database for Astrochemistry \citep{mce2013} proposes the neutral--neutral gas-phase reaction \mbox{C$_2$H$_2$ + Si → SiC$_2$ + H$_2$} for the formation of SiC$_2$. To compute the rate constant, $k(T)$, we use the expression: 
\begin{equation}
k(T) = \alpha \left(  \frac{T}{300}  \right)^\beta \rm exp(-\gamma/T)
,\end{equation} 
 where $\alpha=1.3\times10^{-10}$ cm$^3$ s$^{-1}$, $\beta$ = $-$ 0.71, and $\gamma=29$ K based on experiments by \citet{can2001}; see also UMIST. Then, the calculated $k(T)$ at typical $T_{\rm kin}$ in GC clouds, that is, 70$-$150 K, is of the order $10^{-10}$  s$^{-1}$. If atomic silicon is released during the disruption of dust grains, then a reaction with C$_2$H$_2$ is an effective possible route for the formation of SiC$_2$ in the gas phase. According to mid-infrared observations, C$_2$H$_2$ is one of the most abundant molecules in the inner regions of circumstellar envelopes where dust formation is expected, exhibiting a fractional abundance of $8\times10^{-5}$ \citep{fon2008}. This molecule is also regarded as being key to the formation of dust nucleation clusters, such as polycyclic aromatic hydrocarbons, and later to grain growth through addition of molecules to the surface \citep{che1992}.

\section{Conclusion} \label{section:conclusion}
We searched for Si-bearing species toward G\,+0.693, a chemically rich molecular cloud located within the Sgr B2 complex in the GC. The cloud is believed to be affected by shock waves driven by cloud--cloud collisions in the region. We report the first detection of SiC$_2$ toward G\,+0.693 through six rotational transitions. We obtained a fractional abundance of several 10$^{-11}$ relative to H$_2$ for SiC$_2$. We also detect eight rotational transitions of SiS and four transitions of Si$^{18}$O toward the same cloud. The derived fractional abundances are a few 10$^{-10}$ for SiS and 10$^{-9}$ for SiO, which is in agreement with previous observational studies. 

We find that the fraction of silicon locked in SiC$_2$, SiS, and SiO in G\,+0.693 is low compared to that found in the circumstellar envelopes of AGB stars. We investigate the origin of SiC$_2$ emission in the ISM and conclude that SiC$_2$ can be formed in the gas phase by a reaction between the sputtered atomic silicon and C$_2$H$_2$, or it can be released directly from the dust grains due to disruption.

 Our conclusions on the formation of SiC$_2$ in the ISM are solely based on its detection in G\,+0.693. Further observations in this direction, particularly toward outflows, are needed to better investigate the occurrence and origin of SiC$_2$ in the ISM and to improve our understanding of the interstellar
silicon chemistry. 

\begin{acknowledgements}
The 40m radio telescope at the Yebes Observatory is operated by the Spanish Geographic Institute (IGN, Ministerio de Transportes, Movilidad y Agenda Urbana).
The Yebes 40m observations were carried out through project 20A008. P.d.V. and B.T. thank the support from the Spanish Ministerio de Ciencia e Innovacion (MICIU) through project PID2019-107115GB-C21. B.T. also acknowledges the Spanish MICIU for funding support from grant PID2019-106235GB-I00. We acknowledge financial support through the Spanish grant PID2019-105552RB-C41 funded by MCIN/AEI/10.13039/501100011033.
\end{acknowledgements}

\FloatBarrier
\bibliographystyle{aa} 
\bibliography{mybib} 

\begin{thebibliography}{74}
\expandafter\ifx\csname natexlab\endcsname\relax\def\natexlab#1{#1}\fi

\bibitem[{{Amo-Baladr{\'o}n} {et~al.}(2009){Amo-Baladr{\'o}n},
  {Mart{\'\i}n-Pintado}, {Morris}, {Muno}, \&
  {Rodr{\'\i}guez-Fern{\'a}ndez}}]{amo2009}
{Amo-Baladr{\'o}n}, M.~A., {Mart{\'\i}n-Pintado}, J., {Morris}, M.~R., {Muno},
  M.~P., \& {Rodr{\'\i}guez-Fern{\'a}ndez}, N.~J. 2009, \apj, 694, 943

\bibitem[{{Bordiu} {et~al.}(2022){Bordiu}, {Rizzo}, {Bufano},
  {Quintana-Lacaci}, {Buemi}, {Leto}, {Cavallaro}, {Cerrigone}, {Ingallinera},
  {Loru}, {Riggi}, {Trigilio}, {Umana}, \& {Sciacca}}]{bor2022}
{Bordiu}, C., {Rizzo}, J.~R., {Bufano}, F., {et~al.} 2022, \apjl, 939, L30

\bibitem[{{Canosa} {et~al.}(2001){Canosa}, {Le Picard}, {Gougeon},
  {Rebrion-Rowe}, {Travers}, \& {Rowe}}]{can2001}
{Canosa}, A., {Le Picard}, S.~D., {Gougeon}, S., {et~al.} 2001, \jcp, 115, 6495

\bibitem[{{Castro-Carrizo} {et~al.}(2001){Castro-Carrizo}, {Lucas},
  {Bujarrabal}, {Colomer}, \& {Alcolea}}]{cas2001}
{Castro-Carrizo}, A., {Lucas}, R., {Bujarrabal}, V., {Colomer}, F., \&
  {Alcolea}, J. 2001, \aap, 368, L34

\bibitem[{{Cernicharo} {et~al.}(2017){Cernicharo}, {Ag{\'u}ndez}, {Velilla
  Prieto}, {Gu{\'e}lin}, {Pardo}, {Kahane}, {Marka}, {Kramer}, {Navarro},
  {Quintana-Lacaci}, {Fonfr{\'\i}a}, {Marcelino}, {Tercero}, {Moreno},
  {Massalkhi}, {Santander-Garc{\'\i}a}, {McCarthy}, {Gottlieb}, \&
  {Alonso}}]{cer2017}
{Cernicharo}, J., {Ag{\'u}ndez}, M., {Velilla Prieto}, L., {et~al.} 2017, \aap,
  606, L5

\bibitem[{{Cernicharo} {et~al.}(1989){Cernicharo}, {Gottlieb}, {Guelin},
  {Thaddeus}, \& {Vrtilek}}]{cer1989}
{Cernicharo}, J., {Gottlieb}, C.~A., {Guelin}, M., {Thaddeus}, P., \&
  {Vrtilek}, J.~M. 1989, \apjl, 341, L25

\bibitem[{{Chen} {et~al.}(2022){Chen}, {Xiao}, {Li}, \& {Zhou}}]{che2022}
{Chen}, T., {Xiao}, C.~Y., {Li}, A., \& {Zhou}, C.~T. 2022, \mnras, 509, 5231

\bibitem[{{Cherchneff} {et~al.}(1992){Cherchneff}, {Barker}, \&
  {Tielens}}]{che1992}
{Cherchneff}, I., {Barker}, J.~R., \& {Tielens}, A. G.~G.~M. 1992, \apj, 401,
  269

\bibitem[{{Dickinson} \& {Kuiper}(1981)}]{dic1981}
{Dickinson}, D.~F. \& {Kuiper}, E.~N.~R. 1981, \apj, 247, 112

\bibitem[{{Fonfr{\'\i}a} {et~al.}(2008){Fonfr{\'\i}a}, {Cernicharo}, {Richter},
  \& {Lacy}}]{fon2008}
{Fonfr{\'\i}a}, J.~P., {Cernicharo}, J., {Richter}, M.~J., \& {Lacy}, J.~H.
  2008, \apj, 673, 445

\bibitem[{{Goldsmith} \& {Langer}(1999)}]{gol1999}
{Goldsmith}, P.~F. \& {Langer}, W.~D. 1999, \apj, 517, 209

\bibitem[{{Guesten} \& {Henkel}(1983)}]{gue1983}
{Guesten}, R. \& {Henkel}, C. 1983, \aap, 125, 136

\bibitem[{{G{\"u}sten} \& {Philipp}(2004)}]{gus2004}
{G{\"u}sten}, R. \& {Philipp}, S.~D. 2004, in The Dense Interstellar Medium in
  Galaxies, ed. S.~{Pfalzner}, C.~{Kramer}, C.~{Staubmeier}, \&
  A.~{Heithausen}, Vol.~91, 253

\bibitem[{{Hackwell}(1972)}]{hac1972}
{Hackwell}, J.~A. 1972, \aap, 21, 239

\bibitem[{{Huettemeister} {et~al.}(1998){Huettemeister}, {Dahmen},
  {Mauersberger}, {Henkel}, {Wilson}, \& {Martin-Pintado}}]{hut1998}
{Huettemeister}, S., {Dahmen}, G., {Mauersberger}, R., {et~al.} 1998, \aap,
  334, 646

\bibitem[{{Jim{\'e}nez-Serra} {et~al.}(2008){Jim{\'e}nez-Serra}, {Caselli},
  {Mart{\'\i}n-Pintado}, \& {Hartquist}}]{jim2008}
{Jim{\'e}nez-Serra}, I., {Caselli}, P., {Mart{\'\i}n-Pintado}, J., \&
  {Hartquist}, T.~W. 2008, \aap, 482, 549

\bibitem[{{Jim{\'e}nez-Serra} {et~al.}(2020){Jim{\'e}nez-Serra},
  {Mart{\'\i}n-Pintado}, {Rivilla}, {Rodr{\'\i}guez-Almeida}, {Alonso Alonso},
  {Zeng}, {Cocinero}, {Mart{\'\i}n}, {Requena-Torres}, {Mart{\'\i}n-Domenech},
  \& {Testi}}]{jim2020}
{Jim{\'e}nez-Serra}, I., {Mart{\'\i}n-Pintado}, J., {Rivilla}, V.~M., {et~al.}
  2020, Astrobiology, 20, 1048

\bibitem[{{Jim{\'e}nez-Serra} {et~al.}(2005){Jim{\'e}nez-Serra},
  {Mart{\'\i}n-Pintado}, {Rodr{\'\i}guez-Franco}, \& {Mart{\'\i}n}}]{jim2005}
{Jim{\'e}nez-Serra}, I., {Mart{\'\i}n-Pintado}, J., {Rodr{\'\i}guez-Franco},
  A., \& {Mart{\'\i}n}, S. 2005, \apjl, 627, L121

\bibitem[{{Jim{\'e}nez-Serra} {et~al.}(2022){Jim{\'e}nez-Serra},
  {Rodr{\'\i}guez-Almeida}, {Mart{\'\i}n-Pintado}, {Rivilla}, {Melosso},
  {Zeng}, {Colzi}, {Kawashima}, {Hirota}, {Puzzarini}, {Tercero}, {de Vicente},
  {Rico-Villas}, {Requena-Torres}, \& {Mart{\'\i}n}}]{jim2022}
{Jim{\'e}nez-Serra}, I., {Rodr{\'\i}guez-Almeida}, L.~F.,
  {Mart{\'\i}n-Pintado}, J., {et~al.} 2022, \aap, 663, A181

\bibitem[{{Koeppen} \& {Kegel}(1980)}]{koe1980}
{Koeppen}, J. \& {Kegel}, W.~H. 1980, \aaps, 42, 59

\bibitem[{{Langer} \& {Glassgold}(1990)}]{lan1990}
{Langer}, W.~D. \& {Glassgold}, A.~E. 1990, \apj, 352, 123

\bibitem[{Lefloch {et~al.}(2018)Lefloch, Bachiller, Ceccarelli, Cernicharo,
  Codella, Fuente, Kahane, López-Sepulcre, Tafalla, Vastel, Caux,
  González-García, Bianchi, Gómez-Ruiz, Holdship, Mendoza, Ospina-Zamudio,
  Podio, Quénard, Roueff, Sakai, Viti, Yamamoto, Yoshida, Favre, Monfredini,
  Quitián-Lara, Marcelino, Boechat-Roberty, \& Cabrit}]{asai2018}
Lefloch, B., Bachiller, R., Ceccarelli, C., {et~al.} 2018, Monthly Notices of
  the Royal Astronomical Society, 477, 4792

\bibitem[{{Liechti} \& {Wilson}(1996)}]{lie1996}
{Liechti}, S. \& {Wilson}, T.~L. 1996, \aap, 314, 615

\bibitem[{{Lodders}(2003)}]{lod2003}
{Lodders}, K. 2003, \apj, 591, 1220

\bibitem[{{Mancini} {et~al.}(2022){Mancini}, {Trinari}, {Valen{\c{c}}a Ferreira
  de Arag{\~a}o}, {Rosi}, \& {Balucani}}]{man2022}
{Mancini}, L., {Trinari}, M., {Valen{\c{c}}a Ferreira de Arag{\~a}o}, E.,
  {Rosi}, M., \& {Balucani}, N. 2022, arXiv e-prints, arXiv:2212.11754

\bibitem[{{Mart{\'\i}n} {et~al.}(2019){Mart{\'\i}n}, {Mart{\'\i}n-Pintado},
  {Blanco-S{\'a}nchez}, {Rivilla}, {Rodr{\'\i}guez-Franco}, \&
  {Rico-Villas}}]{mar2019}
{Mart{\'\i}n}, S., {Mart{\'\i}n-Pintado}, J., {Blanco-S{\'a}nchez}, C.,
  {et~al.} 2019, \aap, 631, A159

\bibitem[{{Mart{\'\i}n} {et~al.}(2008){Mart{\'\i}n}, {Requena-Torres},
  {Mart{\'\i}n-Pintado}, \& {Mauersberger}}]{mar2008}
{Mart{\'\i}n}, S., {Requena-Torres}, M.~A., {Mart{\'\i}n-Pintado}, J., \&
  {Mauersberger}, R. 2008, \apj, 678, 245

\bibitem[{{Martin-Pintado} {et~al.}(1992){Martin-Pintado}, {Bachiller}, \&
  {Fuente}}]{mar1992}
{Martin-Pintado}, J., {Bachiller}, R., \& {Fuente}, A. 1992, \aap, 254, 315

\bibitem[{{Mart{\'\i}n-Pintado} {et~al.}(1997){Mart{\'\i}n-Pintado}, {de
  Vicente}, {Fuente}, \& {Planesas}}]{mar1997}
{Mart{\'\i}n-Pintado}, J., {de Vicente}, P., {Fuente}, A., \& {Planesas}, P.
  1997, \apjl, 482, L45

\bibitem[{{Massalkhi} {et~al.}(2019){Massalkhi}, {Ag{\'u}ndez}, \&
  {Cernicharo}}]{mas2019}
{Massalkhi}, S., {Ag{\'u}ndez}, M., \& {Cernicharo}, J. 2019, \aap, 628, A62

\bibitem[{{Massalkhi} {et~al.}(2020){Massalkhi}, {Ag{\'u}ndez}, {Cernicharo},
  \& {Velilla-Prieto}}]{mas2020}
{Massalkhi}, S., {Ag{\'u}ndez}, M., {Cernicharo}, J., \& {Velilla-Prieto}, L.
  2020, \aap, 641, A57

\bibitem[{{Massalkhi} {et~al.}(2018){Massalkhi}, {Ag{\'u}ndez}, {Cernicharo},
  {Velilla Prieto}, {Goicoechea}, {Quintana-Lacaci}, {Fonfr{\'\i}a}, {Alcolea},
  \& {Bujarrabal}}]{mas2018}
{Massalkhi}, S., {Ag{\'u}ndez}, M., {Cernicharo}, J., {et~al.} 2018, \aap, 611,
  A29

\bibitem[{{McElroy} {et~al.}(2013){McElroy}, {Walsh}, {Markwick}, {Cordiner},
  {Smith}, \& {Millar}}]{mce2013}
{McElroy}, D., {Walsh}, C., {Markwick}, A.~J., {et~al.} 2013, \aap, 550, A36

\bibitem[{Minh {et~al.}(2015)Minh, Liu, Su, Hsieh, Liu, Kim, Wright, \&
  Ho}]{min2015}
Minh, Y.~C., Liu, H.~B., Su, Y.-N., {et~al.} 2015, The Astrophysical Journal,
  808, 86

\bibitem[{{Morris} {et~al.}(1975){Morris}, {Gilmore}, {Palmer}, {Turner}, \&
  {Zuckerman}}]{mor1975}
{Morris}, M., {Gilmore}, W., {Palmer}, P., {Turner}, B.~E., \& {Zuckerman}, B.
  1975, \apjl, 199, L47

\bibitem[{{M{\"u}ller} {et~al.}(2005){M{\"u}ller}, {Schl{\"o}der}, {Stutzki},
  \& {Winnewisser}}]{mul2005}
{M{\"u}ller}, H. S.~P., {Schl{\"o}der}, F., {Stutzki}, J., \& {Winnewisser}, G.
  2005, Journal of Molecular Structure, 742, 215

\bibitem[{Pickett {et~al.}(1998)Pickett, Poynter, Cohen, Delitsky, Pearson, \&
  M{\"u}ller}]{pic1998}
Pickett, H., Poynter, R., Cohen, E., {et~al.} 1998, Journal of Quantitative
  Spectroscopy and Radiative Transfer, 60, 883

\bibitem[{{Podio} {et~al.}(2017){Podio}, {Codella}, {Lefloch}, {Balucani},
  {Ceccarelli}, {Bachiller}, {Benedettini}, {Cernicharo}, {Faginas-Lago},
  {Fontani}, {Gusdorf}, \& {Rosi}}]{pod2017}
{Podio}, L., {Codella}, C., {Lefloch}, B., {et~al.} 2017, \mnras, 470, L16

\bibitem[{{Requena-Torres} {et~al.}(2008){Requena-Torres},
  {Mart{\'\i}n-Pintado}, {Mart{\'\i}n}, \& {Morris}}]{req2008}
{Requena-Torres}, M.~A., {Mart{\'\i}n-Pintado}, J., {Mart{\'\i}n}, S., \&
  {Morris}, M.~R. 2008, \apj, 672, 352

\bibitem[{{Requena-Torres} {et~al.}(2006){Requena-Torres},
  {Mart{\'\i}n-Pintado}, {Rodr{\'\i}guez-Franco}, {Mart{\'\i}n},
  {Rodr{\'\i}guez-Fern{\'a}ndez}, \& {de Vicente}}]{req2006}
{Requena-Torres}, M.~A., {Mart{\'\i}n-Pintado}, J., {Rodr{\'\i}guez-Franco},
  A., {et~al.} 2006, \aap, 455, 971

\bibitem[{{Rivilla} {et~al.}(2022{\natexlab{a}}){Rivilla}, {Colzi},
  {Jim{\'e}nez-Serra}, {Mart{\'\i}n-Pintado}, {Meg{\'\i}as}, {Melosso},
  {Bizzocchi}, {L{\'o}pez-Gallifa}, {Mart{\'\i}nez-Henares}, {Massalkhi},
  {Tercero}, {de Vicente}, {Guillemin}, {Garc{\'\i}a de la Concepci{\'o}n},
  {Rico-Villas}, {Zeng}, {Mart{\'\i}n}, {Requena-Torres}, {Tonolo},
  {Alessandrini}, {Dore}, {Barone}, \& {Puzzarini}}]{riv2022a}
{Rivilla}, V.~M., {Colzi}, L., {Jim{\'e}nez-Serra}, I., {et~al.}
  2022{\natexlab{a}}, \apjl, 929, L11

\bibitem[{{Rivilla} {et~al.}(2022{\natexlab{b}}){Rivilla}, {Garc{\'\i}a De La
  Concepci{\'o}n}, {Jim{\'e}nez-Serra}, {Mart{\'\i}n-Pintado}, {Colzi},
  {Tercero}, {Meg{\'\i}as}, {L{\'o}pez-Gallifa}, {Mart{\'\i}nez-Henares},
  {Massalkhi}, {Mart{\'\i}n}, {Zeng}, {De Vicente}, {Rico-Villas},
  {Requena-Torres}, \& {Cosentino}}]{riv2022b}
{Rivilla}, V.~M., {Garc{\'\i}a De La Concepci{\'o}n}, J., {Jim{\'e}nez-Serra},
  I., {et~al.} 2022{\natexlab{b}}, Frontiers in Astronomy and Space Sciences,
  9, 829288

\bibitem[{{Rivilla} {et~al.}(2021{\natexlab{a}}){Rivilla}, {Jim{\'e}nez-Serra},
  {Garc{\'\i}a de la Concepci{\'o}n}, {Mart{\'\i}n-Pintado}, {Colzi},
  {Rodr{\'\i}guez-Almeida}, {Tercero}, {Rico-Villas}, {Zeng}, {Mart{\'\i}n},
  {Requena-Torres}, \& {de Vicente}}]{riv2021a}
{Rivilla}, V.~M., {Jim{\'e}nez-Serra}, I., {Garc{\'\i}a de la Concepci{\'o}n},
  J., {et~al.} 2021{\natexlab{a}}, \mnras, 506, L79

\bibitem[{{Rivilla} {et~al.}(2021{\natexlab{b}}){Rivilla}, {Jim{\'e}nez-Serra},
  {Mart{\'\i}n-Pintado}, {Briones}, {Rodr{\'\i}guez-Almeida}, {Rico-Villas},
  {Tercero}, {Zeng}, {Colzi}, {de Vicente}, {Mart{\'\i}n}, \&
  {Requena-Torres}}]{riv2021b}
{Rivilla}, V.~M., {Jim{\'e}nez-Serra}, I., {Mart{\'\i}n-Pintado}, J., {et~al.}
  2021{\natexlab{b}}, Proceedings of the National Academy of Science, 118

\bibitem[{{Rivilla} {et~al.}(2022{\natexlab{c}}){Rivilla}, {Jim{\'e}nez-Serra},
  {Mart{\'\i}n-Pintado}, {Colzi}, {Tercero}, {de Vicente}, {Zeng},
  {Mart{\'\i}n}, {Garc{\'\i}a de la Concepci{\'o}n}, {Bizzocchi}, {Melosso},
  {Rico-Villas}, \& {Requena-Torres}}]{riv2022c}
{Rivilla}, V.~M., {Jim{\'e}nez-Serra}, I., {Mart{\'\i}n-Pintado}, J., {et~al.}
  2022{\natexlab{c}}, Frontiers in Astronomy and Space Sciences, 9, 876870

\bibitem[{{Rivilla} {et~al.}(2020){Rivilla}, {Mart{\'\i}n-Pintado},
  {Jim{\'e}nez-Serra}, {Mart{\'\i}n}, {Rodr{\'\i}guez-Almeida},
  {Requena-Torres}, {Rico-Villas}, {Zeng}, \& {Briones}}]{riv2020}
{Rivilla}, V.~M., {Mart{\'\i}n-Pintado}, J., {Jim{\'e}nez-Serra}, I., {et~al.}
  2020, \apjl, 899, L28

\bibitem[{{Rivilla} {et~al.}(2019){Rivilla}, {Mart{\'\i}n-Pintado},
  {Jim{\'e}nez-Serra}, {Zeng}, {Mart{\'\i}n}, {Armijos-Abenda{\~n}o},
  {Requena-Torres}, {Aladro}, \& {Riquelme}}]{riv2019}
{Rivilla}, V.~M., {Mart{\'\i}n-Pintado}, J., {Jim{\'e}nez-Serra}, I., {et~al.}
  2019, \mnras, 483, L114

\bibitem[{{Rocha} {et~al.}(2022){Rocha}, {Linnartz}, \& {Varandas}}]{roc2022}
{Rocha}, C.~M.~R., {Linnartz}, H., \& {Varandas}, A.~J.~C. 2022, \jcp, 157,
  104301

\bibitem[{{Rodr{\'\i}guez-Almeida}
  {et~al.}(2021{\natexlab{a}}){Rodr{\'\i}guez-Almeida}, {Jim{\'e}nez-Serra},
  {Rivilla}, {Mart{\'\i}n-Pintado}, {Zeng}, {Tercero}, {de Vicente}, {Colzi},
  {Rico-Villas}, {Mart{\'\i}n}, \& {Requena-Torres}}]{rod2021a}
{Rodr{\'\i}guez-Almeida}, L.~F., {Jim{\'e}nez-Serra}, I., {Rivilla}, V.~M.,
  {et~al.} 2021{\natexlab{a}}, \apjl, 912, L11

\bibitem[{{Rodr{\'\i}guez-Almeida}
  {et~al.}(2021{\natexlab{b}}){Rodr{\'\i}guez-Almeida}, {Rivilla},
  {Jim{\'e}nez-Serra}, {Melosso}, {Colzi}, {Zeng}, {Tercero}, {de Vicente},
  {Mart{\'\i}n}, {Requena-Torres}, {Rico-Villas}, \&
  {Mart{\'\i}n-Pintado}}]{rod2021b}
{Rodr{\'\i}guez-Almeida}, L.~F., {Rivilla}, V.~M., {Jim{\'e}nez-Serra}, I.,
  {et~al.} 2021{\natexlab{b}}, \aap, 654, L1

\bibitem[{{Rodr{\'\i}guez-Fern{\'a}ndez}
  {et~al.}(2000){Rodr{\'\i}guez-Fern{\'a}ndez}, {Mart{\'\i}n-Pintado}, {de
  Vicente}, {Fuente}, {H{\"u}ttemeister}, {Wilson}, \& {Kunze}}]{rod2000}
{Rodr{\'\i}guez-Fern{\'a}ndez}, N.~J., {Mart{\'\i}n-Pintado}, J., {de Vicente},
  P., {et~al.} 2000, \aap, 356, 695

\bibitem[{{Rodriguez-Fernandez, N. J.} {et~al.}(2006){Rodriguez-Fernandez, N.
  J.}, {Combes, F.}, {Martin-Pintado, J.}, {Wilson, T. L.}, \& {Apponi,
  A.}}]{rod2006}
{Rodriguez-Fernandez, N. J.}, {Combes, F.}, {Martin-Pintado, J.}, {Wilson, T.
  L.}, \& {Apponi, A.} 2006, A\&A, 455, 963

\bibitem[{{Rosi} {et~al.}(2018){Rosi}, {Mancini}, {Skouteris}, {Ceccarelli},
  {Faginas Lago}, {Podio}, {Codella}, {Lefloch}, \& {Balucani}}]{ros2018}
{Rosi}, M., {Mancini}, L., {Skouteris}, D., {et~al.} 2018, Chemical Physics
  Letters, 695, 87

\bibitem[{{Roveri} {et~al.}(1988){Roveri}, {Erdelyi Mendes}, \&
  {Singh}}]{rov1988}
{Roveri}, R.~M., {Erdelyi Mendes}, M., \& {Singh}, P.~D. 1988, \aap, 199, 127

\bibitem[{{Schilke} {et~al.}(1997){Schilke}, {Groesbeck}, {Blake}, {Phillips},
  \& {T.~G.}}]{sch1997}
{Schilke}, P., {Groesbeck}, T.~D., {Blake}, G.~A., {Phillips}, \& {T.~G.} 1997,
  \apjs, 108, 301

\bibitem[{{Schilke} {et~al.}(2003){Schilke}, {Leurini}, {Menten}, \&
  {Alcolea}}]{sch2003}
{Schilke}, P., {Leurini}, S., {Menten}, K.~M., \& {Alcolea}, J. 2003, \aap,
  412, L15

\bibitem[{{Sch{\"o}ier} {et~al.}(2007){Sch{\"o}ier}, {Bast}, {Olofsson}, \&
  {Lindqvist}}]{sch2007}
{Sch{\"o}ier}, F.~L., {Bast}, J., {Olofsson}, H., \& {Lindqvist}, M. 2007,
  \aap, 473, 871

\bibitem[{{Sch{\"o}ier} {et~al.}(2006){Sch{\"o}ier}, {Olofsson}, \&
  {Lundgren}}]{sch2006}
{Sch{\"o}ier}, F.~L., {Olofsson}, H., \& {Lundgren}, A.~A. 2006, \aap, 454, 247

\bibitem[{{Tercero} {et~al.}(2011){Tercero}, {Vincent}, {Cernicharo}, {Viti},
  \& {Marcelino}}]{ter2011}
{Tercero}, B., {Vincent}, L., {Cernicharo}, J., {Viti}, S., \& {Marcelino}, N.
  2011, \aap, 528, A26

\bibitem[{{Tercero} {et~al.}(2021){Tercero}, {L{\'o}pez-P{\'e}rez}, {Gallego},
  {Beltr{\'a}n}, {Garc{\'\i}a}, {Patino-Esteban}, {L{\'o}pez-Fern{\'a}ndez},
  {G{\'o}mez-Molina}, {Diez}, {Garc{\'\i}a-Carre{\~n}o}, {Malo}, {Amils},
  {Serna}, {Albo}, {Hern{\'a}ndez}, {Vaquero}, {Gonz{\'a}lez-Garc{\'\i}a},
  {Barbas}, {L{\'o}pez-Fern{\'a}ndez}, {Bujarrabal}, {G{\'o}mez-Garrido},
  {Pardo}, {Santander-Garc{\'\i}a}, {Tercero}, {Cernicharo}, \& {de
  Vicente}}]{ter2021}
{Tercero}, F., {L{\'o}pez-P{\'e}rez}, J.~A., {Gallego}, J.~D., {et~al.} 2021,
  \aap, 645, A37

\bibitem[{{Thaddeus} {et~al.}(1984){Thaddeus}, {Cummins}, \& {Linke}}]{tha1984}
{Thaddeus}, P., {Cummins}, S.~E., \& {Linke}, R.~A. 1984, \apjl, 283, L45

\bibitem[{{Tsuji}(1973)}]{tsu1973}
{Tsuji}, T. 1973, \aap, 23, 411

\bibitem[{{Turner}(1992)}]{tur1992}
{Turner}, B.~E. 1992, \apjl, 388, L35

\bibitem[{{van der Tak} {et~al.}(2007){van der Tak}, {Black}, {Sch{\"o}ier},
  {Jansen}, \& {van Dishoeck}}]{van2007}
{van der Tak}, F.~F.~S., {Black}, J.~H., {Sch{\"o}ier}, F.~L., {Jansen}, D.~J.,
  \& {van Dishoeck}, E.~F. 2007, \aap, 468, 627

\bibitem[{{Wilson} {et~al.}(1971){Wilson}, {Penzias}, {Jefferts}, {Kutner}, \&
  {Thaddeus}}]{wil1971}
{Wilson}, R.~W., {Penzias}, A.~A., {Jefferts}, K.~B., {Kutner}, M., \&
  {Thaddeus}, P. 1971, \apjl, 167, L97

\bibitem[{{Wilson}(1999)}]{wil1999}
{Wilson}, T.~L. 1999, Reports on Progress in Physics, 62, 143

\bibitem[{{Wilson} \& {Rood}(1994)}]{wil1994}
{Wilson}, T.~L. \& {Rood}, R. 1994, \araa, 32, 191

\bibitem[{{Wolff}(1980)}]{wol1980}
{Wolff}, R.~S. 1980, \apj, 242, 1005

\bibitem[{{Zanchet} {et~al.}(2018){Zanchet}, {Roncero}, {Ag{\'u}ndez}, \&
  {Cernicharo}}]{zan2018}
{Zanchet}, A., {Roncero}, O., {Ag{\'u}ndez}, M., \& {Cernicharo}, J. 2018,
  \apj, 862, 38

\bibitem[{{Zeng} {et~al.}(2018){Zeng}, {Jim{\'e}nez-Serra}, {Rivilla},
  {Mart{\'\i}n}, {Mart{\'\i}n-Pintado}, {Requena-Torres},
  {Armijos-Abenda{\~n}o}, {Riquelme}, \& {Aladro}}]{zen2018}
{Zeng}, S., {Jim{\'e}nez-Serra}, I., {Rivilla}, V.~M., {et~al.} 2018, \mnras,
  478, 2962

\bibitem[{{Zeng} {et~al.}(2021){Zeng}, {Jim{\'e}nez-Serra}, {Rivilla},
  {Mart{\'\i}n-Pintado}, {Rodr{\'\i}guez-Almeida}, {Tercero}, {de Vicente},
  {Rico-Villas}, {Colzi}, {Mart{\'\i}n}, \& {Requena-Torres}}]{zen2021}
{Zeng}, S., {Jim{\'e}nez-Serra}, I., {Rivilla}, V.~M., {et~al.} 2021, \apjl,
  920, L27

\bibitem[{{Zeng} {et~al.}(2020){Zeng}, {Zhang}, {Jim{\'e}nez-Serra}, {Tercero},
  {Lu}, {Mart{\'\i}n-Pintado}, {de Vicente}, {Rivilla}, \& {Li}}]{zen2020}
{Zeng}, S., {Zhang}, Q., {Jim{\'e}nez-Serra}, I., {et~al.} 2020, \mnras, 497,
  4896

\bibitem[{{Ziurys}(1988)}]{ziu1988}
{Ziurys}, L.~M. 1988, \apj, 324, 544

\bibitem[{{Ziurys}(1991)}]{ziu1991}
{Ziurys}, L.~M. 1991, \apj, 379, 260

\end{thebibliography}
\end{document}